\documentstyle[11pt] {article}
\newfont{\ssf}{cmss10 scaled 1000}
\def\al{\alpha}

\def\ga{\gamma}
\def\Q{{\cal Q}}

\def\an{\at^{-\nu}}
\def\at{\left| t \right|}
\def\atom{\frac{\an}{M}}
\def\an{\at^{-\nu}}
\def\ba{\begin{eqnarray}}
\def\cee{\chi_{1,1}}
\def\cm{\chi_{M}}

\def\cmfm{\chi^{mf}_{M}}
\def\de{\Delta_{1}}
\def\dst{\displaystyle}
\def\ea{\end{eqnarray}}
\def\eqtzero{\setcounter{equation}{0}}

\def\gee{\ga_{1,1}}
\def\H{{\cal H}}
\def\he{{h_{1}}}

\def\Hn{{\cal H}_{0}}

\def\lotm{\log \tm^{-1}}
\def\lot{\log\at^{-1}}
\def\lb{\left(}
\def\lcb{\left\{}
\def\lg{\langle}

\def\nn{\nonumber}

\def\phe{\prt \he}

\def\pzfm{\prt^{\,2} \!f_{M}}
\def\prt{\partial}
\def\pzt{\prt^{\,2}}
\def\pr{\prime}
\def\prop{\propto}

\def\Qe{\Q_{1}}
\def\Qh{\Q_{hom}}
\def\Qz{\Q_{2}}
\def\ra{\rightarrow}
\def\rb{\right)}
\def\rcb{\right\}}
\def\rg{\rangle}
\def\rgn{\rangle_{0}}
\def\sst{\scriptscriptstyle}
\def\su{susceptibility}
\def\susz{\su}
\def\tc{T_{c}}
\def\TM{T_{M}}
\def\tm{{t_{\sst M}}}
\def\tmm{\tm^{-\nu}/M}
\def\tmmf{\frac{\tm^{-\nu}}{M}}

\def\zm{z_{\sst M}}

\begin{document}
\title {\bf
Finite--Size Scaling Analysis of Generalized Mean--Field Theories
\\{~}\\}

\author{
  Steffen D.~Frischat\thanks{Present address: Max--Planck--Institut
f\"ur Kernphysik, Postfach 103980, 69029 Heidelberg, FRG}\mbox{ } and Reimer
K\"uhn\thanks{Supported by a Heisenberg fellowship}
}

\date
      {
          Institut f\"ur Theoretische Physik   \\
                Universit\"at Heidelberg \\
                    Philosophenweg 19\\
             69120 Heidelberg, Germany\\
      }

\maketitle

  \vspace {3ex}

\begin{abstract}\noindent
We investigate families of generalized mean--field theories that can
be formulated using the Peierls--Bogoliubov inequality. For
test--Hamiltonians describing mutually non--interacting subsystems of
increasing size,  the thermodynamics of these mean--field type systems
approaches that of the infinite, fully interacting system except in
the immediate vicinity of their respective mean--field critical
points. Finite--size scaling analysis of this mean--field critical
behaviour allows to extract the critical exponents of the fully
interacting system. It turns out that this procedure amounts to the
coherent anomaly method (CAM) proposed by Suzuki, which is thus given
a transparent interpretation in terms of conventional renormalization
group ideas. Moreover, given the geometry of approximating systems, we can
identify the family of approximants which is optimal in the sense of the
Peierls--Bogoliubov inequality. In the case of the 2--$d$ Ising model
it turns out that, surprisingly, this optimal family gives rise to a
spurious singularity of thermodynamic functions.
\end{abstract}

\section{Introduction}
Standard wisdom has it that closed form approximations and
renormalization group methods play complementary roles in the analysis
of thermodynamic behaviour of many--particle systems. The former
usually generate mean--field type theories, and as such often
provide efficient tools to obtain a good {\it qualitative} picture of
a given system's thermodynamics. Equations of state and qualitatively
correct phase diagrams are relatively easily calculated. Famous
examples of such approaches are the van der Waals theory of imperfect
gases and the Weiss self--consistent theory of ferromagnetism. With
respect to a {\it quantitative} description of phase transitions,
however, these theories invariably fail and produce wrong critical
exponents. Renormalization group ideas, on the other hand, provide a
satisfactory theoretical description of critical phenomena and the
interplay of critical exponents. Except for the determination of
critical exponents, though, renormalization group calculations are
rather involved and do not easily allow to obtain a picture of the system's
thermodynamic properties.

In the course of time, various refinements of the standard mean--field
theory have been proposed; for an overview, see \cite{Bur 72}. Patterned after
the first such attempt due to Bethe \cite{Bet 35}, short range
correlations of the
dynamic variables are taken into account by considering small
clusters. Equations of state are generated in the form of {\it
self--consistency} equations which impose certain physically plausible
constraints, such as homogenity of the order parameter. This inclusion
of short range correlations, hence of additional phase space, leads to
improved (i.~e., lower) estimates of the critical temperature. But it
fails to produce improved critical exponents -- the reason being that
critical phenomena are dominated by {\it long--range} correlations.

An alternative, more systematic construction of mean--field type
theories derives from a {\it variational} scheme based on the
Peierls--Bogoliubov (PB) inequality \cite{Pei 34}. This inequality is based on
convexity arguments and states that, given a system with Hamiltonian $\H$, its
free energy $F$ can be approximated from above by the trial ``free
energy'' $\Phi$ as
\ba
 F \leq \Phi := F_{0} + \lg \H - \Hn   \rg_{0} = \lg \H \rg_0 - T S_0\ .
\label{peierls} \ea

Here, $\Hn$ is an arbitrary test--Hamiltonian for the system in question, which
depends on some set $\{h_\alpha\}$ of variational parameters. $\lg \dots \rg_0$
denotes the average over the Gibbs distribution generated by $\Hn$,
and $F_0$ and $S_0$ denote the corresponding free energy and entropy.
The idea is to choose
$\Hn$ such that the corresponding Gibbs distribution is analytically or
numerically tractable and to determine the variational parameters $h_\alpha$
so as to minimize the right hand side of (\ref{peierls}). The resulting
minimization conditions replace the above self--consistency equations
and generate the system's equations of state.
To state an example, let $\H$ describe the Ising spin system on a
lattice in $d$ space dimensions. The simplest approximating $\Hn$ then
describes a system of non--interacting spins in a mean field $h_0$,
that is, $\Hn =-h_0 \sum_i S_i$. Minimizing the corresponding trial
free energy with respect to $h_0$ generates the conventional Weiss
mean--field equation of state. For a recent application of this
method to CsNiF${}_3$ chains, see \cite{Trud}.

Short range correlations can now be taken into account by choosing a
system of mutually independent {\it clusters} of spins, which together
make up the whole system. Increasing the size of these clusters, one
obtains a scheme of approximations that should systematically approach
the thermodynamics of the underlying, fully interacting (spin) system.

Despite the fact that every PB system of finite (or quasi one--dimensional)
geometry exhibits mean--field type critical behaviour, the true critical
exponents of the underlying system can be extracted by invoking
finite--size scaling (FSS) ideas. Thermodynamic functions can be
evaluated as functions of cluster size. We will show that this
procedure is equivalent to Suzuki's coherent anomaly method (CAM)
\cite{Suz 86, Suz 87}. CAM is thus demonstrated to be firmly rooted in
the FSS philosophy and hence in conventional renormalization group ideas.

Within the general PB scheme, and for a given cluster geometry, various
families of approximating systems can be constructed which differ in
number, symmetries, or even nature of their variational parameters. Of
all of those, the optimal family -- in the sense of minimal trial free
energy -- is the one with the largest set of independent variational
parameters compatible with the symmetries of the system.

In the present paper, we will explore a collection of approximating
sequences for the 2--$d$ Ising model. Two families will prove to be of
special interest: cyclically ``closed'' strips, that display Suzuki's
coherent anomaly, and ``open'' strips of lower symmetry, that can be
identified as the optimal PB sequence. Surprisingly, this optimal
family gives rise to a spurious singularity of thermodynamic
functions making any extrapolation to the full 2--$d$ model based on the open
strip's mean--field critical behaviour impossible. In this
restricted sense, the
complementarity between closed form approximations and the RG
reappears.

The outline of our paper is as follows. In Sec.~2, we introduce a collection of
variational trial systems for the 2--$d$ Ising system, based on the
Peierls--Bogoliubov
inequality. Test--Hamiltonians will be defined on $M\times\infty$ strips which
may be open or closed in the $M$ direction. We show that, quite generally, all
but one of the extremization conditions for the variational parameters
can be solved explicitly, leaving only one non--trivial condition as
the equation of state. In Sec.~3, we use finite--size scaling to
derive the scaling of the mean--field critical temperatures $T_M$ with
strip width $M$. For test--ensemles based on cyclically closed strips
(Sec.~3.1), this provides a first method to extract the susceptibility
exponent $\gamma$ of the underlying, fully interacting system. The
second method uses the mean--field susceptibilities and their scaling
with strip width $M$. The coherent anomaly, which is at the base of
this method, is given a simple explanation as a standard FSS
phenomenon. Sec~3.2 is devoted to an analysis of variational
approximants defined on open strips. These were identified in Sec.~2
as the best sequence of variational approximants of strip--geometry in
the framework of the PB inequality. Contrary to expectations, a FSS
analysis of these ``optimal'' variational approximants predicts a
spurious singularity of thermodynamic functions that precludes any
extrapolation attempt to the two--dimensional system. Sec.~4 is a
discussion of the results.

\section{Variational approximants for the 2--$d$ Ising model}
\eqtzero

We now introduce a collection of approximating systems
for the 2--$d$ Ising model, based on the Peierls--Bogoliubov inequality. Let
\ba
\H = - J \sum_{(ij)} s_i s_j - H \sum_i s_i
\ea
describe the fully interacting system on a square lattice of $N'\times N$
Ising spins, and let us assume periodic boundary conditions in both directions.
As a test--ensemble we use a system of mutually {\it
non--interacting\/} strips of
size $M\times N$ with periodic boundary conditions in the ``longitudinal''
$N$ direction, and either free or periodic boundary conditions in the
``transverse''
$M$ direction.

\subsection{Test--ensembles based on closed strips}

Let us first consider the version which is cyclically closed in {\it both\/}
directions. A simple Hamiltonian for a single strip of width $M$,
which exhibits full translational invariance in both directions, is given
by
\ba
\H_{M,N}^c = -h_T\sum_{(ij)_T} s_i s_j - h_L\sum_{(ij)_L} s_i s_j
 - (H+h) \sum_i s_i\ ,
\label{clo:ham}\ea
where we have introduced three variational parameters
$\{h_\al\}:=\{h_T,h_L,h\}$ a coupling $h_T$ for
transverse nearest neighbours $(ij)_T$, a coupling $h_L$ for longitudinal
nearest neighbours $(ij)_L$, and a variational field $h$. The trivial
effect of the external field $H$ has been absorbed into the definition
of $h$.

Let us denote the free energy density of
an isolated strip with Hamiltonian (\ref{clo:ham})by
$f_M^c=f_M^c(T,H,\{h_\al\})$. Assuming $N'$ to
be an integer multiple of $M$, the PB inequality for this setup states that
\ba
F &\leq& \Phi_M^c(T,H,\{h_\al\}) =: N' N\, \phi_M^c(T,H,\{h_\al\})\nn \\[.25cm]
  &=& N' N \left\{ f_M^c + \Big (J \Big (1 - \frac{1}{M}\Big ) - h_T\Big )
\frac{\prt f_M^c}{\prt h_T} + \Big (J-h_L \Big )\frac{\prt f_M^c}{\prt h_L}
- \Big (J\frac{1}{M} \frac{\prt f_M^c}{\prt h} + h \Big )
\frac{\prt f_M^c}{\prt h} \right\}.
\label{clo:PB}\ea
Here we have used the fact that $\langle s_i s_j \rangle_0$ is given by
$(-\prt f_M^c/\prt h_T)$ for transverse and by $(-\prt
f_M^c/\prt h_L)$ for longitudinal nearest neighbours {\it within\/} a strip,
while $\langle s_i s_j \rangle_0 =\langle s_i \rangle_0 \langle s_j
\rangle_0= (-\prt f_M^c / \prt h)^2 =: (m_M^c)^2 $ for spins belonging to
{\it different\/} strips. Here $m_M^c$ denotes the magnetization of a
strip of width $M$.

{}From (\ref{clo:PB}) the minimization conditions are obtained in the form
\ba
\varphi_\alpha := \frac{\prt\phi_M^c}{\prt h_\alpha} = \sum_\beta \frac{\prt^2
f_M^c}{\prt h_\alpha \prt h_\beta} \psi_\beta = 0
\label{clo:vec}\ea
with $h_\alpha \in \{h_T,h_L,h\}$, and
\ba
\{\:\psi_\beta\:\} =  \lcb J \Big (1 - \frac{1}{M}\Big ) - h_T\ ,\
J - h_L\ ,\ J \frac{2}{M} m_M^c - h \rcb \ .
\label{clo:psi}\ea
Due to the concavity of $f_M^c$ as a function of the $h_\alpha$ -- the Hessian
$(\prt^2 f_M^c/\prt h_\alpha \prt h_\beta)_{\al,\beta}$ is a strictly negative
definite matrix -- the solution of (\ref{clo:vec}) can be read off
immediately: it is $\psi_\beta = 0$, i.~e.,
\ba
h_T &=& J \Big(1 - \frac{1}{M}\Big) \label{clo:ht}\\
h_L &=& J \label{clo:hl}\\
h &=& J \frac{2}{M} m_M^c =- J \frac{2}{M} \frac{\prt f_M^c}{\prt h} \ .
\label{clo:sce}
\ea
Of these, only the last one, (\ref{clo:sce}), is a non--trivial
transcendental equation
with a solution that varies with temperature $T$ and external field
$H$. It determines the mean field $h=h(T,H)$. The variational
(mean--field) free energy density computed within
this approach is then
\ba
f_M^{mf}(T,H) = \phi_M^c(T,H,h_T,h_L,h(T,H))\ ,
\label{clo:fmmf}\ea
with variational parameters $\{h_\al\}$ determined by (\ref{clo:ht}) --
(\ref{clo:sce}).

Thermodynamic functions are obtained by differentiation of $f_M^{mf}(T,H)$
along the solution manifold given by (\ref{clo:sce}). This yields
\ba
m_M^{mf} = - \left(\frac{d f_M^{mf}(T,H)}{d H}\right )_{\varphi,T} =
-\frac{\prt \phi_M^c}{\prt H} -\sum_\alpha \frac{\prt \phi_M^c}{\prt h_\alpha}
\left(\frac{d h_\alpha}{d H}\right )_{\varphi,T}\ ,
\ea
where the subscripts $\varphi,T$ denote differentiation along the manifold
$\varphi_\alpha = 0$ at constant $T$. Partial differentials are taken
as usual. With the help of
(\ref{clo:fmmf}) and $\prt \phi_M^c /\prt h_\alpha \equiv 0$ one obtains the
mean--field magnetization
\ba
m_M^{mf} = - \frac{\prt f_M^c}{\prt H}\ .
\ea
In a similar vein, the mean--field susceptibility is found to be
\ba
\chi_M^{mf} &=& \left(\frac{d m_M^{mf}(T,H)}{d H}\right )_{\varphi,T}
=\frac{\prt m_M^{mf}}{\prt H} +\sum_\alpha \frac{\prt m_M^{mf}}{\prt h_\alpha}
\left(\frac{d h_\alpha}{d H}\right )_{\varphi,T} \nn \\
 &=& -\frac{\prt^2 f_M^c}{\prt H^2} + J \frac{2}{M}\,\frac{\left( \,\prt^2
f_M^c/\prt H\prt h\,\right)^2} {\: 1 + J \frac{2}{M}\:\prt^2
f_M^c/\prt h^2\:} \ .
\label{clo:susz}\ea
That is, magnetization and mean--field susceptibility can be expressed in
terms of {\it free} partial derivatives of the strip free energy $f_M^c$ of a
strip; the strip being described by the Hamiltonian $\H_{MN}^c$
evaluated at parameter
values given by (\ref{clo:ht}) -- (\ref{clo:sce}).

In principle, one may try to improve the approximation by introducing
additional variational parameters that represent ``generalized'' couplings
beyond the ones already contained in (\ref{clo:ham})
which generate
interaction terms added to $\H_{MN}^c$ in a translationally invariant
way. To be specific, we modify $\H_{M,N}^c$ according to
\ba
\H_{M,N}^c \longrightarrow \H_{M,N}^c - \sum_{\omega\subseteq\Omega} h_\omega
\sum_i \left(\prod_{j\in\omega} s_{j+i}\right)\ ,
\label{morepar}\ea
where $\Omega$ denotes a collection of subsets of the $M\times N$ strip which
are mutually non--equivalent under translation. It turns out that such an
enlarged space of variational parameters does {\it not\/} actually improve the
variational free energy, because the enlarged set of minimization conditions is
solved by (\ref{clo:ht}) -- (\ref{clo:sce}) and $h_\omega = 0$ for all
of the added $\omega \subseteq \Omega$. To see this,
note that the modification (\ref{morepar}) implies a corresponding replacement
\ba
\lg \Hn \rgn \longrightarrow \lg \Hn \rgn +
N N^{\pr} \sum_{\omega \subseteq \Omega} h_{\omega} \frac{\prt f_M^c}
{\prt h_\omega}\ ,
\ea
where $f_M^c$ is now the free energy corresponding to the modified Hamiltonian
(\ref{morepar}). Hence the enlarged set of minimization conditions can be
formulated in complete analogy to (\ref{clo:vec}), albeit with an
enlarged set of variational parameters, $h_\alpha\in \{h_T , h_L , h
,\{h_\omega\}_{\omega \subseteq \Omega}\}$ and
\ba
\lcb\:\psi_\beta\:\rcb =  \lcb J \Big (1 - \frac{1}{M}\Big ) - h_T\ ,\
J - h_L\ ,\ J\frac{2}{M} m_M^c - h\ ,\ \lcb -h_\omega\rcb_{\omega
\subseteq \Omega} \rcb\ .
\ea
Due to the concavity of $f_M^c(T,H,\{h_\alpha\})$ the assertion follows, that
is $h_\omega = 0$ for all $\omega \subseteq \Omega$.

\subsection{Test--ensembles based on open strips}

An alternative sequence of test systems is defined by considering
``open'' $M\times
N$ strips with free boundary conditions in the transverse $M$
direction. While such strips retain the full translational invariance
in the closed $N$ direction, they exhibit only a reflection symmetry $j
\to M+1-j$ in the open $M$ direction. This reduced symmetry group
allows to introduce a
considerably larger set of independent variational parameters. A
simple Hamiltonian
respecting these symmetries is given by
\ba
\H_{MN}^o & = &  - \sum_{i=1}^{N}\Bigg\{\sum_{j=1}^{\mu}\: h_{T,j}
\sum_{\kappa\in \{j,M-j\}}
s_{i,\kappa} s_{i,\kappa + 1}
\:+\: \sum_{j=1}^{\mu^\pr} \:h_{L,j} \sum_{\kappa\in \{j,M+1-j\}}
s_{i,\kappa} s_{i+1,\kappa}\nn \\
 & & \qquad + \sum_{j=1}^{\mu^\pr} \:(H+ h_j) \sum_{\kappa\in \{j,M+1-j\}}
s_{i,\kappa}\ \Bigg\} \ ,
\label{open:ham}\ea
where $\mu= [M/2]$ and $\mu^\pr=[(M+1)/2]$ with the convention that $[k]$
denotes the largest integer less than or equal to $k$.
Also, we have introduced a two dimensional notation to label the
vertices of the strip. Note that the variational fields and couplings vary from
row
to row, but respect the reflection invariance of the open strip in the $M$
direction. The total number of independent variational parameters  is
$3M/2$ for even $M$, and $(3M+1)/2$ for odd $M$.

Denoting by $f_M^o=f_M^o(T,H,\{h_{T,j}\},\{h_{L,j}\},\{h_j\})$ the free energy
density of an isolated (open) strip with Hamiltonian (\ref{open:ham}),
and assuming $N'$ to be an integer multiple of $M$,
we conclude by the Peierls--Bogoliubov inequality that
\ba
F &\leq& \Phi_M^o(T,H,\{h_{T,j}\},\{h_{L,j}\},\{h_j\}) \nn \\[.25cm]
  &=& N' N \Bigg\{ f_M^o + \sum_{j=1}^{\mu} (J - h_{T,j})
M \frac{\prt f_M^o}{\prt h_{T,j}}
+ \sum_{j=1}^{\mu^\pr} (J - h_{L,j}) M \frac{\prt f_M^o}{\prt h_{L,j}}\nn \\
& & \qquad - \sum_{j=2}^{\mu^\pr} h_{j} M \frac{\prt f_M^o}{\prt h_{j}}
- \Big (J \frac{M}{2} \frac{\prt f_M^o}{\prt h_1} + h_1 \Big )
M \frac{\prt f_M^o}{\prt h_1} \Bigg\} \ .
\ea
The minimization conditions are formally the
same as (\ref{clo:vec}),
with $f_M^c$ replaced by $f_M^o$, with $h_\alpha \in \{ \{h_{T,j}\} ,
\{h_{L,j}\} , \{h_j\} \}$, and
\ba
\{\:\psi_\beta\:\} =  \lcb \{J - h_{T,j}\}\ ,\
\{J - h_{L,j}\}\ ,\ \{-h_j\}_{j=2,..,\mu^\pr}\ ,\ J m_{M,1}^o - h_1
\rcb \ .
\ea
Here $m_{M,1}^o = \langle s_{i,1} \rangle_o = - (M/2) \,  \prt
f_M^o/\prt h_1$. Again, due to concavity, the solution of the minimization
conditions are $\psi_\beta = 0$, or
\ba
h_{T,j} &=& J \label{open:htj} \ ,\\
h_{L,j} &=& J \ , \\
h_{j} &=& 0 \qquad\qquad {\rm for}\qquad j = 2,\dots , \mu^\pr \ , \\
h_1 &=& J m_{M,1}^o = - J \frac{M}{2} \frac{\prt f_M^o}{\prt h_1} \ .
\label{open:sce}
\ea
That is, all variational couplings are equal to the coupling $J$ of the
underlying system, and all variational fields, except for the
boundary field $\he$, vanish.

Thermodynamic functions are obtained as before. In particular, the
mean--field--magnetization of an ``open'' strip of width $M$ is given by
\ba
m_M^{mf} = - \frac{\prt f_M^o}{\prt H}\ ,
\ea
and the mean--field \susz\ by
\ba
\chi_M^{mf} = -\frac{\prt^2 f_M^o}{\prt H^2} + J \frac{M}{2}\,
\frac{\left(\,\prt^2
f_M^o/\prt H\prt h_1\,\right )^2} {\:1 + J \frac{M}{2}\: \prt^2
f_M^o/\prt h_1^2} \ .
\label{open:susz}\ea
As above, thermodynamic functions can be expressed in terms of {\it free}
partial derivatives of the free energy of a single strip of corresponding
geometry described by the Hamiltonian (\ref{open:ham}) with parameter
values given by (\ref{open:htj}) --
(\ref{open:sce}).

As in the case of closed approximants, any attempt to enlarge the space
of variational parameters  by adding further (multi--spin) interactions to
$\H_{MN}^o$ does not lead to any improvement of the variational
approximations: the minimization condition requires the
corresponding coupling constants to be zero. In particular, an extra
variational
coupling $h_{T,M}$ which would {\it close\/} the strip in the
transverse $M$ direction will have to vanish, rendering the
strip open again at optimally chosen variational parameters. Therefore,
within the framework of strip geometries, the test--ensemble based on
open strips with Hamiltonian (\ref{open:ham}) may be identified as
{\it optimal\/} in the sense of the PB inequality. It uses the largest
meaningful set of independent variational parameters compatible with
the lowest symmetry of $M\times\infty$ strips. Hence, the minimum
obtained by $f_M^o$ is the total minimum of sensible trial free
energies $\Phi$.

\section{Finite--size scaling analysis of variational approximants}
\eqtzero

We now turn to an evaluation of thermodynamic functions
computed within the variational approximation schemes described in Sec.~2.
The dependence of the thermodynamic behaviour of mean--field
test--strips on their width $M$ will be extracted by the use of
finite--size scaling. Wherever possible, we will determine critical exponents.

In both cases, only one variational parameter turned out to be
non--trivial. In the case of test--ensembles living on closed strips,
this parameter is a variational field $h$ acting {\it homogeneously\/} on
all spins and determined by (\ref{clo:sce}),
\ba
h = J \frac{2}{M} m_M^c\ ; \nn
\ea
whereas in the case of test--ensembles based on open strips, it is a
{\it boundary} field $h_1$ acting only on the first and the $M$--th
row of each strip, and which is determined by (\ref{open:sce}),
\ba
h_1 = J  m_{M,1}^o\ . \nn
\ea
In both versions of the variational scheme, the approximation $T_M$ of the
critical point $T_c$ is signalled by the appearance of non--zero solutions
of the variational field, $h$ or $h_1$, respectively. As the temperature $T$
is lowered, the bifurcation from the zero solution occurs when
(setting $J=1$)
\ba
1= \frac{2}{M} \frac{\prt m_M^c}{\prt h}(T,H=0,h=0)
\label{critc}
\ea
in the ``closed'' variant, and when
\ba
1=  \frac{\prt m_{M,1}^o}{\prt h_1}(T,H=0,h_1=0)
\label{crito}
\ea
in the ``open'' variant. The solutions of these equations define the
mean--field critical temperatures $T_M$. In the following, we discuss
the scaling analysis of
these equations and of the corresponding divergence of the mean--field
susceptibilities (\ref{clo:susz}) and (\ref{open:susz}). The two
different setups will be considered seperately.

\subsection{Finite--size analysis of test--ensembles based on closed strips}

Let us begin with the sequence of approximants based on closed homogeneously
magnetized strips. In this setup both the variational field $h$ and
the external field $H$ act homogeneously on all spins. As
we have chosen $\H_{MN}^c$ to depend on these fields through their sum
$H+h$, the free energy $f_M^c$, is a function of $(H+h)$ only, and we can
replace partial
derivatives of $f_M^c$ with respect to $h$ by  partial derivatives
with respect to $H$ and
vice versa. Denoting the ``free'' \su\ of a closed strip of
circumference $M$ by
\ba
\chi_M^c(T,h) = -\frac{\prt^2 f_M^c}{\prt H^2}(T,H=0,h) \ ,
\ea
 we can reformulate the critical condition (\ref{critc}) as
\ba
1=\frac{2}{M}\chi_M^c(\TM,0) \ ,
\label{critconhom}\ea
and the expression (\ref{clo:susz}) for the mean--field susceptibility
according to
\ba
\cmfm(T,H=0) &=& \cm^c(T,h) + \frac{2}{M} \frac{\lb\cm^c(T,h)\rb^2}
{1-\frac{2}{M}
\cm^c(T,h) }\nn \\
&=& \frac{\cm^c(T,h)}{1-\frac{2}{M}\cm^c(T,h)} \;.
\label{suszhom}\ea
These two expressions are now directly amenable to analysis by FSS
\cite{Bar 83}.

Analysis  of (\ref{critconhom}) will give the asymptotic behaviour of the
reduced mean--field critical temperatures $\tm := (\TM -\tc)/\TM$.
Let us first recall the finite--size behaviour of the free energy of an Ising
strip of width $M$ in zero field. Close to the critical temperature $\tc$ of
the two dimensional Ising system, the singular part of its zero field \su\ is
given by the finite--size scaling expression
\ba
\cm(T)\sim \chi_{\infty}(T)\,\Q(\frac{\xi_{\infty}(T)}{M}) \sim
\at^{-\ga}\Qh(\frac{\at^{-\nu}}{M})\ ,
\label{suszscal}\ea
where $t = (T-T_c)/T_c$. This expression holds for open and closed strips
alike, albeit with different scaling functions.
The behaviour of the scaling function $\Qh(z)$ in the limits $z:=(
\an/M) \ra 0$ and $z \ra \infty$ can be easily determined. These limits
correspond to the cases $M \ra \infty$ at non--critical
temperature $T\neq \tc$, or $\at \ra 0$ at finite size $M < \infty$,
respectively. Regularity of the left hand side of (\ref{suszscal}) in
these cases
implies the power laws $\Qh(z) \sim 1$ for $z \ra 0$ and $\Qh(z) \sim
z^{-\ga/\nu}$ for $z \ra\infty$.

This can be applied to the strips considered above. At
temperatures above and at the mean--field critical temperature $\TM$,
no variational field is present. Under the assumption that the
temperatures $\TM$ are sufficiently
close to $\tc$ for large $M$, standard finite--size scaling holds,
and (\ref{critconhom}) becomes
\ba
1 \sim \frac{\tm^{-\gamma}}{M}\Qh(\tmmf) \ .
\label{critscalhom}
\ea
As $M\to \infty$, the argument
$\zm:=(\tmm)$ of the
scaling function at $T=T_M$ can either vanish, converge to a non--zero
constant, or diverge. The latter two cases lead to contradictions
($\gamma \neq \nu$
assumed), leaving $\zm\ra 0$ as the only possibility. Hence $\Qh(\zm)
\sim 1$ as $M\ra\infty$, and by (\ref{critscalhom}) the mean--field
critical temperatures
$\TM$ asymptotically scale as
\ba
\tm = (\TM -\tc)/\tc \sim M^{-1/\ga}\;.
\label{TMhom} \ea
Note that $\gamma$ is the true susceptibility exponent of the underlying 2--$d$
Ising model so that (\ref{TMhom}) can be used to determine both $T_c$ and
$\gamma$.

The same analysis, applied to (\ref{suszhom}), gives the behaviour of the
mean--field susceptibility $\cmfm$ in the vicinity of the mean--field critical
temperatures $T_M$. Expanding the denominator in small temperature
differences $t-\tm=(T-\TM)/\tc$ above $\TM$ gives
\ba
1-\frac{2}{M}t^{-\gamma}\Qh(z) \simeq (t-\tm)\frac{2}{M} \:
\tm^{-\ga-1}\left[ \ga \Qh(\zm) + \nu\zm\Qh^{\pr}(\zm)\right] \ .
\ea
The second term in the square brackets can
be neglected, as $\zm \ra 0$ and $\Qh^{\pr} \ra 0$ for $M\ra\infty$.
We substitute $M \prop \tm^{-\ga}$, cancel $\cm^c(\TM,0) \prop
\tm^{-\ga}\Qh(\zm)$, and finally arrive at
\ba
\cmfm(T)\; \prop \; \frac{1}{(t - \tm)}\: \frac{1}{\tm^{\ga -1}}\;.
\label{homcam} \ea
Eq. (\ref{homcam}) exhibits the usual mean--field divergence of the
susceptibility $\cmfm(T)\; \prop \; (t - \tm)^{-1}$ as $t\to\tm$. Note
that the prefactor $\tm^{-(\ga -1)}$ itself diverges as
$M\to\infty$, a phenomenon
for which Suzuki coined the term {\it coherent anomaly} \cite{Suz 86}.
Obviously, the coherent
anomaly provides a second possibility to extract the asymptotic susceptibility
exponent $\gamma$ of the underlying system from the sequence of mean--field
approximations. Our considerations clearly show that this method is entirely
rooted in the FSS philosophy, hence in conventional renormalization
group ideas. This relationship has hitherto been much less clear in
the literature.

The dependence of the mean--field critical temperatures $T_M$ on the strip
width $M$ is displayed in Fig.~1. The convergence to the asymptotic value is
fairly slow, as can be anticipated from (3.8). Nevertheless, good extrapolation
algorithms (see, for instance, \cite{Henkel}) predict $T_M\to T_\infty =
2.26\pm 0.01$, which is reasonably close
to the exact value of $T_c \simeq 2.269$. Setting $T_c = T_\infty$ in
(3.8), we
obtain an estimate for $\gamma$, viz., $\gamma \to 1.77\pm 0.03$ as $M\to
\infty$. While not extraordinary, this result is also not too far off the mark.

Fig.~2 shows the values of $\gamma$, obtained from the ratio of
prefactors $\bar\chi_M=\tm^{-(\ga -1)}$ of the mean--field susceptibility
(3.10) for two successive strip widths $M$ and $M'=M+1$. Assuming
$T_c=T_\infty$, we
extrapolate this sequence of $\gamma$ values to $\gamma_\infty = 1.765\pm
0.01$, which produces a reasonably good agreement with the exact
result $\gamma=1.75$ for the susceptibility exponent. With the exact
value for
$T_c$, the extrapolation yields a slightly better result, $\gamma_\infty =
1.751\pm 0.01$, which gives the susceptibility exponent to within less
than $1\%$
of the exact result.

\subsection{Finite--size analysis of test--ensembles based on open strips}

Much to our surprise, the approximation scheme breaks down in the
case of the ensemble of open, inhomogeneously magnetized strips -- the
family of systems we identified as ideal in the sense of the
Peierls--Bogoliubov inequality!

By concavity, we have singled out the boundary field $\he$ as the only
non--trivial variational parameter. It affects only two spins per
column. The self--consistency equation is given by (\ref{open:sce}),
and the critical condition by
\ba
1= \frac{\prt m_{M,1}^o}{\phe}(\TM,0,0)\ .
\label{critconinh}\ea
Numerical values of $\TM$, obtained by transfer matrix calculations
are plotted in Fig.~3.
Unexpectedly, as $M\to\infty$, they converge to a temperature
$T_\infty \simeq 2.64$ which is {\it different\/} from the correct
critical temperature $\tc \simeq 2.27$ of the two--dimensional Ising
model.

For a qualitative explanation of this behaviour, we again refer to FSS
analysis. We note that the \su\
\ba
\frac{\prt m_{M,1}^o}{\phe}(T,0,\he)
= - \frac{M}{2}\frac{\pzfm^o}{\prt\he^2}(T,0,\he)=:\cee(T,\he,M)
\ea
that appears in the critical condition (\ref{critconinh}) is well known
in the theory of surface critical phenomena \cite{Bin 83, Bar 83}. It
takes the scaling form
\ba
\cee(T,\he,M)\;\sim\;  \lot \tilde{\Qe}(\he\at^{-\de},\atom) \;+\;
\tilde{\Qz}(\he\at^{-\de},\atom)
\label{ceesimp2}\ea
where $\de$ is the gap exponent corresponding to $\he$ (see \cite{Au-Y
75, Au-Y 80, McC 67} for  extensive treatments). In the limit
$z=(\an/M)\ra0$, which corresponds to first taking the thermodynamic limit
and then approaching the critical temperature, it diverges
logarithmically:
\ba
\cee(T) \,\sim\, \lot.
\label{ceeznull2} \ea
This fact is closely related to the anomalous logarithmic divergence
of the specific heat occuring in the two--dimensional Ising lattice.
Consequently, at finite width and at the critical temperature
($z\ra\infty$), the FSS behaviour of $\cee$ at zero variational field
is given by
\ba
\cee(\tc,M) \; \sim \; \log M \ .
\label{ceezinfty2} \ea
These results at hand, we can now return to the critical condition
\ba
1 \; \sim \; \cee(\TM,0,M) \; \sim \; \lotm \tilde{\Qe}(0,\zm)
\;+\; \tilde{\Qz}(0,\zm) \ .
\label{critconscal2}\ea
Again, the cases $\zm \ra const$ and $\zm\ra\infty$ as $M\to\infty$ can be
ruled out
by (\ref{critconscal2}) and (\ref{ceezinfty2}), respectively. For
$\zm\ra 0$, (\ref{ceeznull2}) becomes $1\sim
\lotm$, or $\tm\sim 1$, which is consistent with $z_M\to 0$. We
therefore conclude that
\ba
\tm =(\TM -\tc)/\tc \,\sim \,1 \qquad (M \ra \infty) \ .
\ea
The critical temperatures of the sequence of test--systems based  on
open inhomogeneously magnetized strips do ${\it not}$ converge to
$\tc$ which conforms to our above observation. Note that this result
is not merely an anomaly of the two--dimensional Ising lattice. It
does not, as it might seem, depend on the logarithmic singularity of
$\cee(T)$. A similar calculation for the case of a non--zero exponent
$\gee$ corresponding to $\cee$ gives just the same behavior for the
$\tm$'s. It has tacitly been assumed, though, that the edge itself
cannot become critical at a temperature different from the bulk's $\tc$.
This restricts the above argument to quasi--one--dimensional
test--systems.

We can finally use these results to investigate the scaling of the mean--field
susceptibilities (\ref{open:susz}) with strip width $M$ near the
respective mean--field
critical points $T_M$. Analogous to (\ref{suszhom}), we find the mean--field
susceptibility
\ba \cmfm(T) \:=\: \cm(T,\he(T))\: +\: J\,\frac{2}{M}\,\frac{\dst \lb
\,\chi_{1}(T,\he(T))\,\rb^{2}}{\; 1-J\cee(T,\he(T))\;} \ .
\label{inhsusz} \ea
Here $\cm = \pzt f/\prt H^{\,2}$, $\chi_{1} = -(2/M)\pzt f/\prt H\prt\he$ and
 $\cee = -(2/M) \pzt f/\prt\he^{2}$ are the ``bulk'' \susz\ and the two
``surface'' susceptibilities well known in the standard treatment of surface
critical phenomena.
The mean--field \susz\ $\cmfm(T)$ differs from the free \susz\ $\cm(T)$ of a
strip of the same geometry in two ways: by the action of the variational
boundary field $\he(T)$ in the first term of the right hand side of
(\ref{inhsusz}), and by an explicit mean--field divergence in the second term.

Both of these contributions can be shown to become irrelevant in the
limit of large $M$. In this limit, $\xi_M(\TM)/M \ra 0$ in view of (3.17),
so that the two surface susceptibilities $\chi_1$ and $\cee$ become independent
of the strip width $M$. Setting them constant and expanding in small
temperature
differences above $\TM$, we arrive at
\ba
\cmfm(T)\,\sim\,\cm(T,\he(T)) + \frac{const}{M}\,\frac{1}{T-\TM} \ .
\ea
Thus, in the limit of large width $M$, the explicit mean--field contribution
is suppressed by a factor $1/M$. Furthermore, the surface field $\he(T)$
appearing below $\TM$ does not affect thermodynamic properties of the bulk at
large $M$, since the thermodynamic limit is {\it independent\/} of boundary
conditions.

We thus encounter the paradoxical situation of a singularity in the ``open''
mean--field approximants which becomes spurious, as the
limit $M\to\infty$ is taken. That is, even though thermodynamic functions
exhibit a (suppressed) singularity at a temperature $T_M > \tc$, this
singularity does {\it not} in the limit $M\to\infty$ correspond to a change in
the system's thermodynamic properties. Evidently, no useful information can be
drawn from these mean--field singularities, and any attempt to extrapolate to
the underlying 2--$d$ Ising lattice must fail.

The true thermodynamic singularity of the mean--field susceptibility develops
right at $T_c$ in the {\it first\/} contribution to (\ref{inhsusz}), by
conventional reduction of the finite--size rounding of the bulk susceptibility
$\chi_M$ as $M\to\infty$.

\section{Conclusions}
\eqtzero

In the present paper a generalized mean--field theory based on
the Peierls--Bogoliubov inequality is used to define
quasi--one--dimensional approximants for the 2--$d$ Ising lattice. By
convexity arguments, all but one of the
minimization conditions for the varational parameters can be solved
explicitly, a finding that is not restricted to the Ising model but
holds generally for systems with a scalar order parameter. Thereby a
systematic classification of PB approximants is possible. We singled
out two types of strips: periodically ``closed'' strips with
rotational symmetry in the direction transverse to the axis of
infinite extent and ``open'' strips of inhomogeneous magnetization.

By standard finite--size scaling, the
former are shown to display a coherent
anomaly. Estimates of critical exponents of the underlying, fully
interacting Ising system can be extracted.
At this point, it has to be stressed that
the variational method presented above should not be advocated as a
new, superior numerical tool for computing transition temperatures or
critical exponents. The estimates calculated above are clearly
inferior to those obtained by Hu {\it et
al.~}\cite{Hu 87}. In the original CAM scheme, based on {\it ad hoc}
self--consistency equations, mean--field critical temperatures behave like
$(\TM^{CAM}-\tc)\sim M^{-1/\nu}$, whereas (\ref{critscalhom}) states
$(\TM^{PB}-\tc)\sim M^{-1/\ga}$. That is, the convergence of the PB
critical temperatures is slower than in the scheme of Hu {\it et al.} (which in
turn is slower than that of the conventional phenomenological
renormalization group procedure \cite{Night}). In the PB scheme,
the asymptotic regime of FSS power laws is reached only for very large
strip width. Corrections to scaling are therefore expected to play an
important r\^ole.

Nevertheless, we believe that our findings are of
interest in their own right, since the appearance of a coherent anomaly in
the family of ``closed'' approximants can be given a transparent
interpretation as a FSS phenomenon.

In sharp contrast to expectations, the ``open'' strips, which are found to be
the optimal family of approximants in the sense of the PB inequality,
give rise to a spurious criticality at a temperature {\it different\,} from
the 2--$d$ Ising critical temperature. Any extrapolation to the full
two--dimensional system based on the mean--field critical behaviour of
this family must therefore fail.  This unexpected result clearly
shows that variational descriptions of many particle systems should be
used with utmost caution.

{\bf Acknowledgements}\\
It is a pleasure to thank Heinz Horner for many stimulating
discussions and Jochen Rau for the careful reading of the manuscript.


\mbox{}\\[2cm]
{\Large\bf Figure Captions}

{\bf Fig. 1:} Mean--field critical temperature $T_M$ of closed approximants
as a function of inverse strip width $1/M$.

{\bf Fig. 2:} Estimate of the susceptibility exponent derived from the ratio
of prefactors $\bar \chi_M$ of the mean-field susceptibility for two
successive strip widths $M$ and $M'=M+1$ as a function of $2/(M+M')$.

{\bf Fig. 3:} Mean-field critical temperature $T_M$ of open approximants
as a function of inverse strip width $1/M$. Note that they do {\it not\/}
extrapolate to $T_c$, as $M\to\infty.$

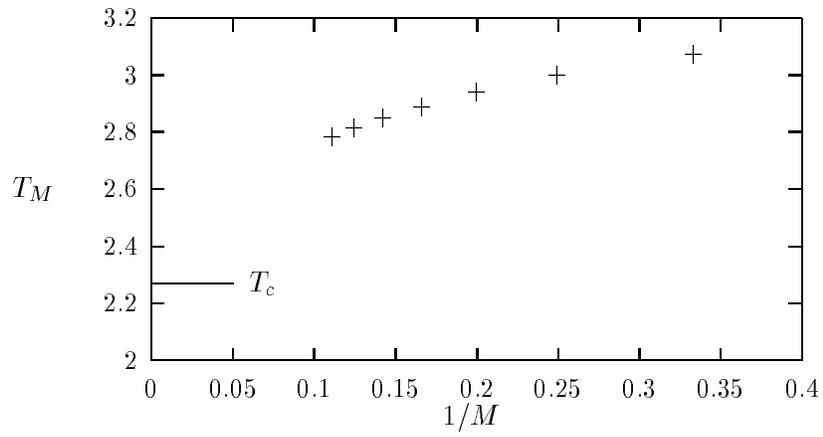
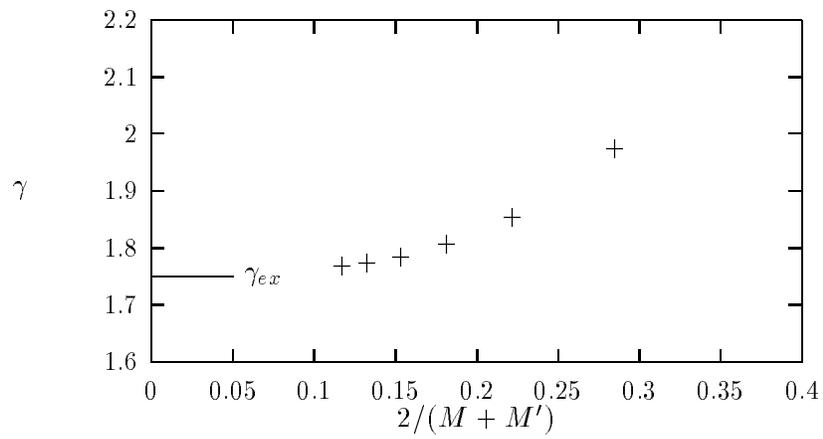
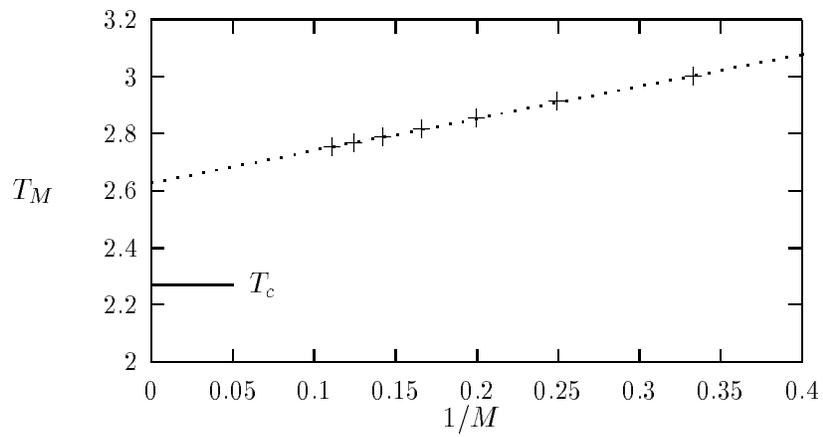
\end{document}